      [1999/12/01 v1.4c Il Nuovo Cimento]
\documentclass{cimento}

\usepackage[
   centertags, 
   sumlimits,  
   intlimits,  
   namelimits, 
]{amsmath} %

\usepackage[final]{graphicx}

\graphicspath{{./pics/}}
\DeclareGraphicsExtensions{.eps}

\title{Collective Flow and Energy Loss with parton transport}

\author{
I.~Bouras\from{ITPFrankfurt}\ETC,
A.~El\from{ITPFrankfurt},
O.~Fochler\from{ITPFrankfurt},
F.~Reining\from{ITPFrankfurt}
J.~Uphoff\from{ITPFrankfurt}
C.~Wesp\from{ITPFrankfurt}\\
Z.~Xu\from{ITPFrankfurt}\from{FIASFrankfurt}
        \atque
C.~Greiner\from{ITPFrankfurt}
}

\instlist{
  \inst{ITPFrankfurt} Institut fuer Theoretische Physik,
		      Goethe Universitaet Frankfurt, Max-von-Laue-Strasse.1,
		      D-60438 Frankfurt am Main, Germany
  \inst{FIASFrankfurt} Frankfurt Institute for Advanced Studies,
		      Ruth-Moufang-Strasse 1,
		      D-60438 Frankfurt am Main, Germany
}

\PACSes
{
\PACSit{25.75.-q}{Relativistic heavy-ion collisions}
\PACSit{25.75.-q}{quark-gluon Plasma}
\PACSit{24.10.Lx}{Monte Carlo simulations}
\PACSit{52.35.Tc}{Shock waves and discontinuities}
\PACSit{25.75.Cj}{heavy quark production in HIC}
}

\begin{document}

\maketitle

\begin{abstract}

Quenching of gluonic jets and heavy quark production
in Au+Au collisions at RHIC can be understood within
the pQCD based 3+1 dimensional
parton transport model BAMPS including pQCD bremsstrahlung
$2 \leftrightarrow 3$ processes. Furthermore, the
development of conical structures induced by gluonic
jets is investigated in a static box for the regimes
of small and large dissipation.

\end{abstract}

\section{Introduction}

The values of the elliptic flow parameter $v_2$ measured by the experiments
at the Relativistic Heavy Ion collider (RHIC)
\cite{Adler:2003kt,Adams:2003am,Back:2004mh} suggest that in the
evolving QCD fireball a fast local equilibration of quarks and gluon occurs
at a very short time scale $\le 1$ fm/c. This locally thermalized state of
matter, the quark gluon plasma (QGP), behaves as a nearly perfect fluid,
confirmed by viscous hydrodynamics \cite{Luzum:2008cw,Song:2008hj} and
microscopic transport theory \cite{Xu:2007jv,Xu:2008av}. The viscosity
to entropy ratio coefficient $\eta/s$
has to be rather small, possibly close to the conjectured lower bound
$\eta/s = 1/4 \pi$ from a correspondence between
conformal field theory and string theory in an Anti-de-Sitter
space \cite{Kovtun:2004de}. To achieve a rather small $\eta/s$ value
in a partonic gas, binary perturbative QCD (pQCD) processes require unphysical
large cross sections and thus inelastic radiative interactions \cite{Xu:2004mz}
become important.

The phenomenon of jet-quenching has been another important discovery
at RHIC \cite{Adams:2003kv}. Hadrons with high transverse momenta are suppressed in
$Au + Au$ collisions with respect to a scaled $p + p$ reference
\cite{Adler:2002xw,Adcox:2001jp}. This quenching of jets is commonly
attributed to energy loss on the partonic level as the hard partons
produced in initial interactions are bound to traverse the QGP created in the 
early stages of heavy-ion collisions (HIC). In addition, very exciting jet-associated particle
correlations have been observed \cite{Wang:2004kfa}, which might be the result of a
conical emission of propagating shock waves in form of Mach Cones induced
by highly-energetic partons traversing the expanding
medium \cite{Stoecker:2004qu}. Furthermore, heavy quarks are interesting
for the investigation of the early stage of the QGP.
Due to their large mass, initially produced heavy quarks,
can cover -- depending on their production point -- a long distance through
the QGP. Interactions on this way and subsequent modifications on heavy quark
distributions can reveal valuable information about the properties of
the medium.

A large class of phenomena in heavy-ion collisions can be investigated
within the framework of the kinetic transport theory. Among others, the
kinetic transport model BAMPS (Boltzmann Approach to Multiparton Scatterings)
\cite{Xu:2004mz} was developed to describe the
early QGP of a HIC. Using BAMPS early thermalization
of gluons within $\tau < 1$ fm/c was demonstrated in Au+Au collisions at
$\sqrt{s_{NN}} = 200$ GeV
employing Glauber initial conditions and the coupling constant $a_s = 0.3$.
In addition to the
elastic pQCD $gg \leftrightarrow gg$ processes, pQCD-inspired
bremsstrahlung $gg \leftrightarrow ggg$ was included. This was shown
to be essential for the achievment of local thermal equilibrium
at that short time scale. The fast thermalization happens also in a
similar way using a Color Glass Condensate as inital conditions
\cite{El:2007vg}.

BAMPS has been applied to simulate elliptic flow and
jet quenching at RHIC energies \cite{Fochler:2008ts} for the first time
using a consistent and fully pQCD--based microscopic transport model to 
approach both key observables on the partonic level within a common setup.
The left panel of Fig. \ref{fig:v2_summary_RAA_central} shows that the
medium simulated in the parton cascade BAMPS exhibits a sizable degree
of elliptic flow in agreement with experimental findings at RHIC
as established in \cite{Xu:2007jv, Xu:2008av}.

The extraction of the shear viscosity over entropy density ratio $\eta/s$
has confirmed the essential importance of inelastic processes.
Within the present description bremsstrahlung and back reaction processes
lower the
shear viscosity to entropy density ratio significantly by a factor of $7$,
compared to the ratio when only elastic collisions are considered
\cite{Xu:2007ns,El:2008yy}. For
$a_s = 0.3$ one finds $\eta/s = 0.13$, where for $a_S = 0.6$ the values
matches the lower bound of $\eta/s = 1/4\pi$ from the AdS/CFT conjecture.

In this paper we demonstrate the description of different phenomena in
relativistic HIC using the relativistic pQCD-based on-shell
parton transport model BAMPS. Due to the large momentum scales
involved the energy loss of partonic jets can be treated in terms of
perturbative QCD (pQCD) and most theoretical schemes attribute the main
contribution to partonic energy loss to radiative processes
\cite{Wicks:2005gt}. In addition, the possible
propagation of Mach Cones in the QGP induced by such highly-energetic
partons can be studied. Considering the earlier work investigating
the effects of dissipation on relativistic shock waves
\cite{Bouras:2009nn,Bouras:2009vs,Bouras:2010hm} we demonstrate
the transition of Mach Cones from ideal to the viscous one.
It is a major challenge to combine jet physics on the one hand and
bulk evolution on the other hand within a common framework. In the
end, the heavy quark production at RHIC and the Large Hadron Collider
(LHC) during the evolution of the QGP is studied.

\section{Jet Quenching in Au+Au collisions at 200 AGeV}

For simulations of Jet Quenching in heavy ion collisions the initial gluon
distributions are sampled according to a mini--jet model with a lower
momentum cut-off $p_{0} = 1.4\,\mathrm{GeV}$ and a $K$--factor of $2$.
The test particle method \cite{Xu:2004mz} is employed to ensure sufficient
statistics.
Quarks are discarded after sampling the initial parton distribution since currently
a purely gluonic medium is considered. To model the freeze out of the
simulated fireball, free streaming is applied to regions where the local
energy density has dropped below a critical energy density
$\varepsilon_{c}$ ($\varepsilon_{c} = 1.0\, \mathrm{GeV}/\mathrm{fm}^3$
unless otherwise noted).

\begin{figure}[tbh]
  \centering
  \begin{minipage}[t]{0.42\textwidth}
    \includegraphics[width=\linewidth]{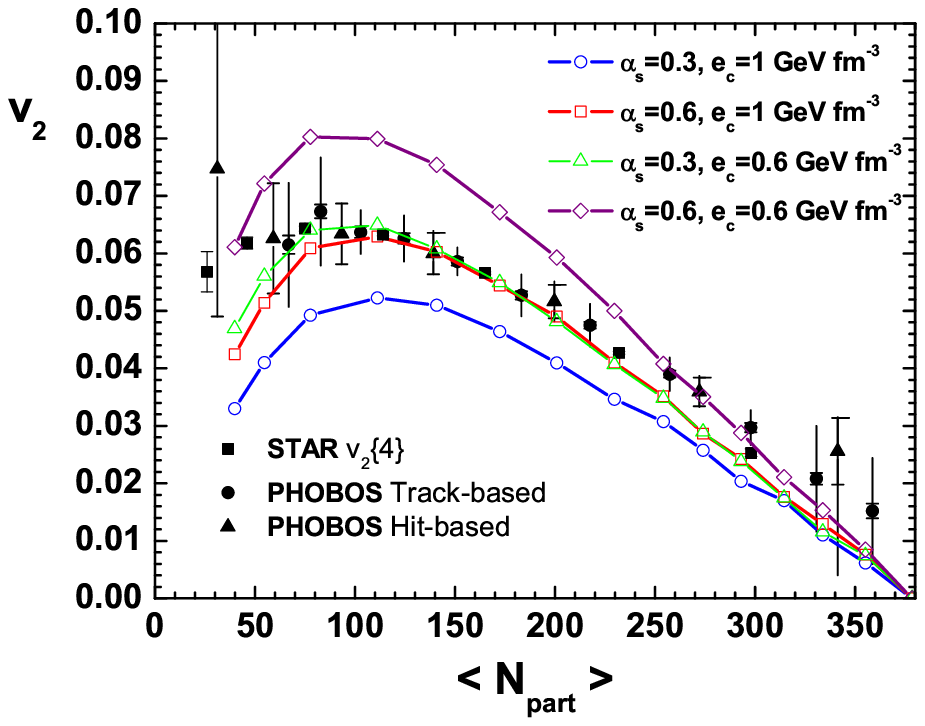}
  \end{minipage}
  \begin{minipage}[t]{0.48\textwidth}
    \includegraphics[width=\linewidth]{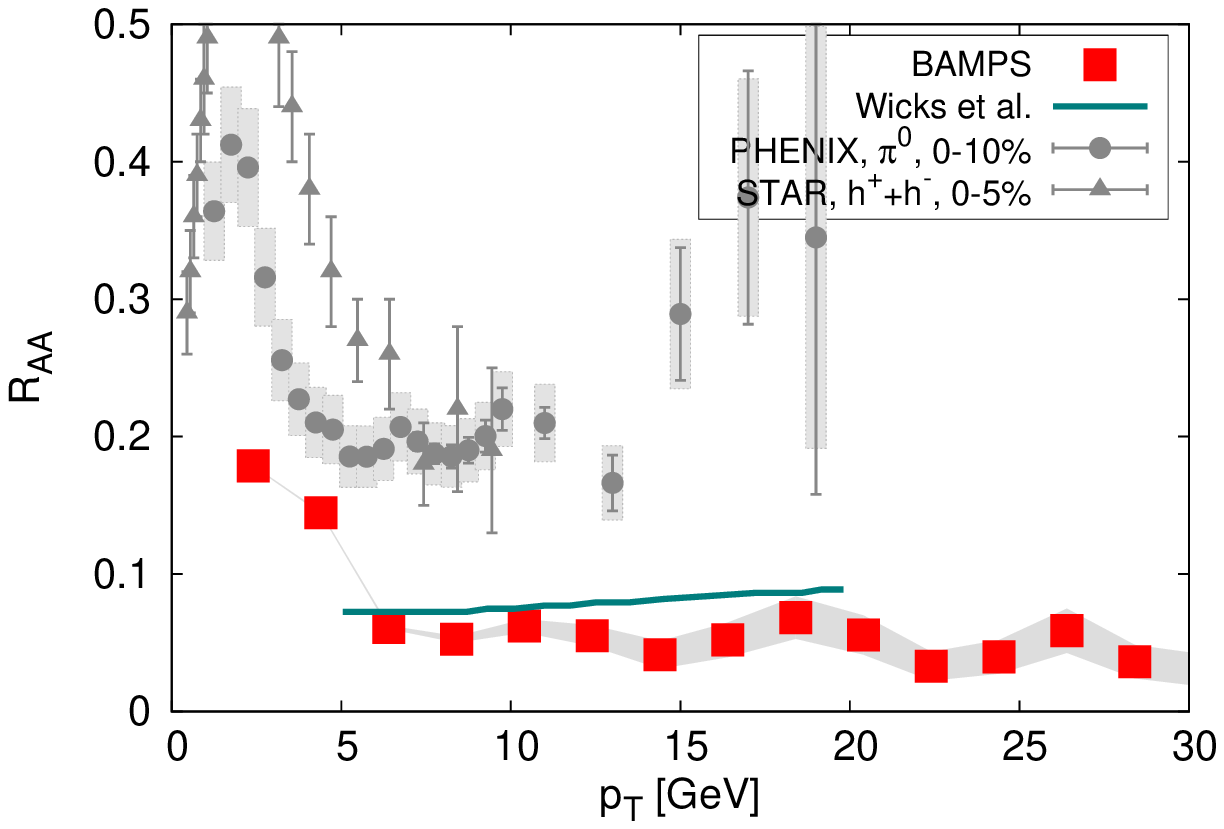}
  \end{minipage}

  \caption{Left panel: Elliptic flow $v_2$ as a function of the number
of participants for Au+Au at 200~AGeV for different combinations of the
strong coupling $\alpha_s$ and the critical energy density $\varepsilon_c$.
See \cite{Xu:2008av} for more information. \newline Right panel: Gluonic
$R_{AA}$ at midrapidity ($y \,\epsilon\, [-0.5,0.5]$) as extracted from
simulations for central Au+Au collisions at 200~AGeV. For comparison the
result from Wicks et al. \cite{Wicks:2005gt} for the gluonic contribution
to $R_{AA}$ and experimental results from PHENIX \cite{Adare:2008qa} for
$\pi^{0}$ and STAR \cite{Adams:2003kv} for charged hadrons are shown.}
\label{fig:v2_summary_RAA_central}
\end{figure}

The right panel of Fig. \ref{fig:v2_summary_RAA_central} shows the gluonic
$R_{AA}$ simulated in BAMPS for central,  $b=0\,\mathrm{fm}$, collisions.
It is roughly constant at $R_{AA}^{\mathrm{gluons}} \approx 0.053$ and in
reasonable agreement with analytic results for the gluonic 
contribution to the nuclear modification factor $R_{AA}$ \cite{Wicks:2005gt},
though the suppression of gluon jets in BAMPS appears to be slightly stronger.
We expect improved agreement in future studies when employing a carefully
averaged $\langle b \rangle$ that will be better suited for comparison to
experimental data than the strict $b=0\,\mathrm{fm}$ case.

\begin{figure}[tbh]
  \centering
  \begin{minipage}[l]{0.45\textwidth}
    \includegraphics[angle=270,width=\linewidth]{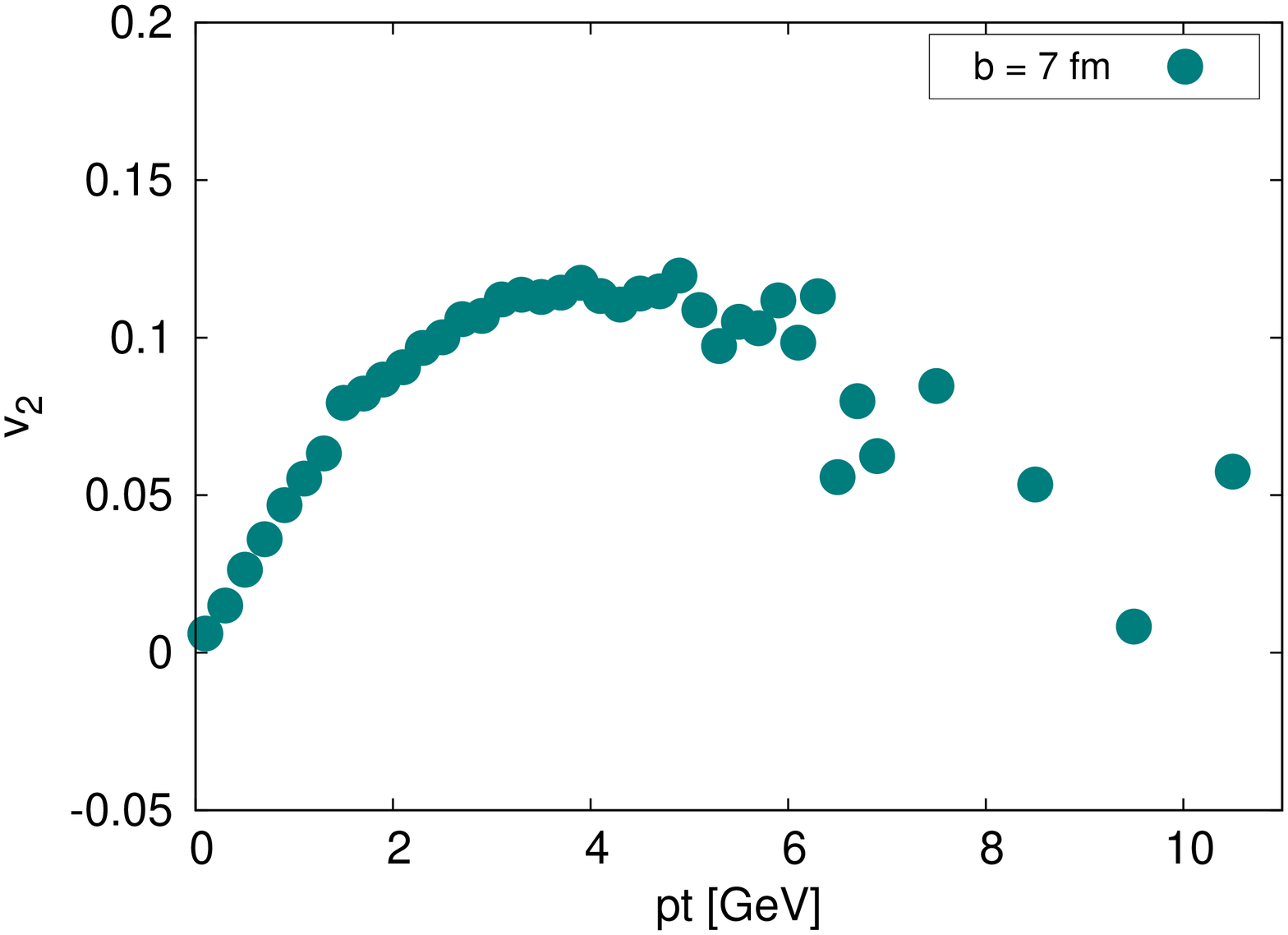}
  \end{minipage}
  \begin{minipage}[r]{0.45\textwidth}
    \includegraphics[width=\linewidth]{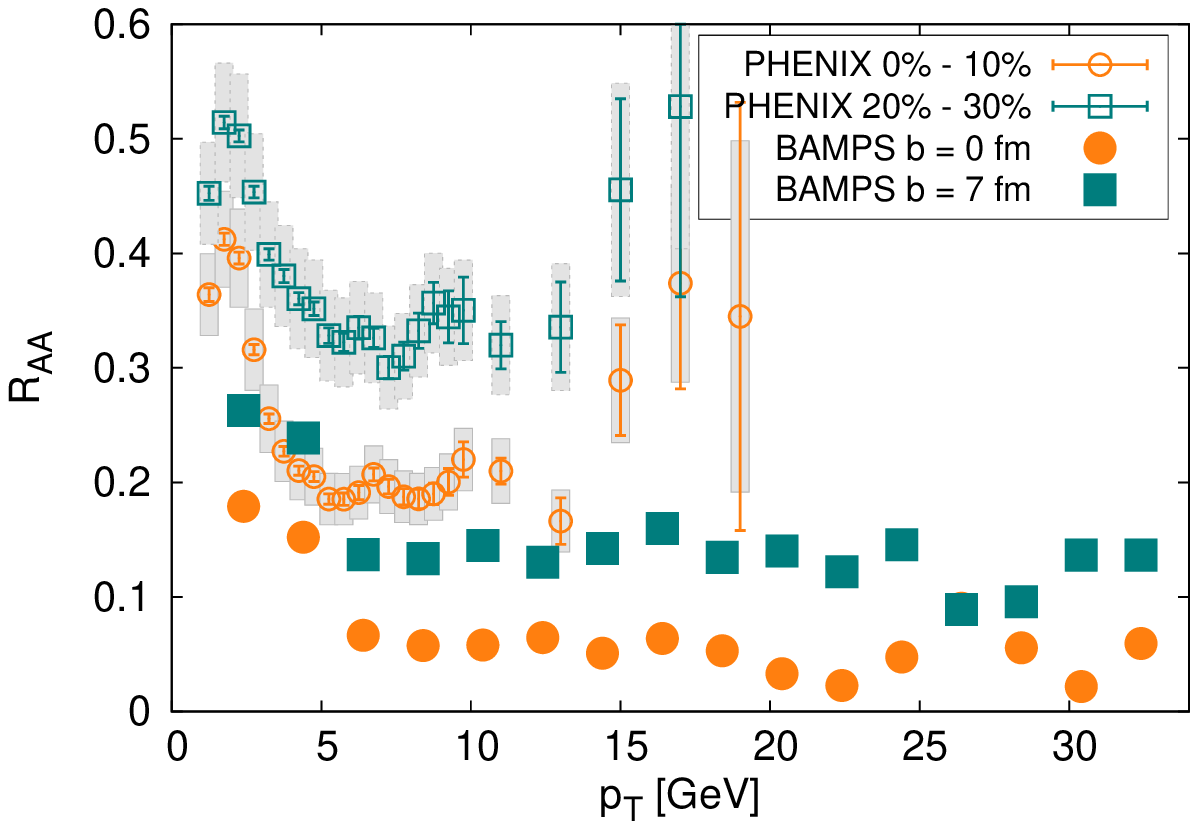}
  \end{minipage}
  \caption{Left panel: Elliptic flow $v_{2}$ for gluons in simulated
Au+Au collisions at 200 AGeV with $b=7\,\mathrm{fm}$. $\varepsilon_c = 0.6\, \mathrm{GeV}/\mathrm{fm}^3$. \newline
  Right panel: Gluonic $R_{AA}$ as extracted from BAMPS simulations for
$b=0\,\mathrm{fm}$ and $b=7\,\mathrm{fm}$, $\varepsilon_c = 1.0\, \mathrm{GeV}/\mathrm{fm}^3$.
For comparison experimental results from PHENIX \cite{Adare:2008qa} for $\pi^{0}$ are shown for
central $(0 - 10) \%$ and off--central $(20-30)\%$ collisions.}
\label{fig:v2_RAA_b7}
\end{figure}

We have computed the gluonic $R_{AA}$ for non--central Au + Au collisions
at the RHIC energy of $\sqrt{s} = 200 \mathrm{AGeV}$ with a fixed impact
parameter $b=7\,\mathrm{fm}$ (Fig. \ref{fig:v2_RAA_b7}), which roughly
corresponds to $(20-30)\%$ experimental centrality. A comparison in
terms of the magnitude of the jet suppression for $b=7\,\mathrm{fm}$ is
difficult since there are no published analytic results available to compare
to. Taking the ratio of the $b=7\,\mathrm{fm}$ to the $b=0\,\mathrm{fm}$
results as a rough guess indicates that the decrease in quenching is more
pronounced in BAMPS compared to experimental data. The ratio of the nuclear
modification factor between central $(0 - 10) \%$ and more peripheral
$(20-30)\%$ collisions is $\left. R_{AA}\right|_{0 \% - 10 \%} / \left. R_{AA}\right|_{20 \% - 30 \%} \approx 0.6$
for the experimental data, while for the BAMPS results
$\left. R_{AA}\right|_{b=0\,\mathrm{fm}} / \left. R_{AA}\right|_{b=7\,\mathrm{fm}} \approx 0.4$.
However, the issue of detailed quantitative comparison needs to be
re-investigated once light quarks and a fragmentation scheme are
included into the simulations.

To complement the investigations of $R_{AA}$ at a non--zero impact parameter
$b=7\,\mathrm{GeV}$, we have computed the elliptic flow parameter $v_{2}$
for gluons at the same impact parameter and extended the range in transverse
momentum up to roughly $p_{T} \approx 10\,\mathrm{GeV}$, see left panel of
Fig. \ref{fig:v2_RAA_b7}. For this a critical energy density
$\varepsilon_{c} = 0.6\, \mathrm{GeV}/\mathrm{fm}^3$ has been used, in order
to be comparable to previous calculations of the elliptic flow within BAMPS.
The $v_{2}$ of high--$p_{T}$ gluons is at first rising with $p_{T}$, but from
$p_{T} \approx 4$ to $5\,\mathrm{GeV}$ on, it begins to slightly decreases again.
This behavior is in good qualitative agreement with recent RHIC data
\cite{Abelev:2008ed} that for charged hadrons shows $v_{2}$ to be rising up
to $v_{2} \approx 0.15$ at $p_{T} \approx 3\,\mathrm{GeV}$ followed by a
slight decrease.


\section{Transition from ideal to dissipative Mach Cones}

There is an important issue in recent studies of
relativistic heavy-ion collisions (HIC) whether the small
but finite viscosity allows the development of
relativistic shocks in form of Mach Cones.
Within the framework of BAMPS studies were finished
to answer the question whether shocks can develop
with finite viscosity and how this will alter such
a picture \cite{Bouras:2009nn}. Within the relativistic
Riemann problem it was shown that one dimensional shocks
smears out if viscosity is large
\cite{Bouras:2009vs,Bouras:2010hm}. However, the expected
viscosity in HIC seems to be small enough to allow a
significant contribution of shocks in form of Mach Cones
into the picture of HIC. In the following we report
a very recent study.

Mach Cones, which are special phenomena of shock waves, have their origin
in ideal hydrodynamics. A very weak perturbation in a
perfect fluid induces sound waves which propagate with the speed
of sound $c_s = \sqrt{dp/de}$, where $p$ is the pressure and $e$ is the
energy density. In the case where the perturbation with velocity $v_{\rm jet}$
propagates faster than the generated sound waves, the sound waves lie on a cone.
Considering a gas of massless particles, where
$e = 3p$ and $c_s = 1/\sqrt{3}$, then the emission angle of
the Mach Cone is given by
$\alpha_w = \arccos ( c_s / v_{\rm jet} ) = 54,73^\circ$.

A stronger perturbation induces the propagation of shock waves
exceeding the speed of sound, therefore the emission angle
changes and can be approximated by\\
$\alpha \approx \arccos ( v_{\rm shock} / v_{\rm jet} )$. Here 
\begin{equation}
\label{eq:v_shock}
v_{\rm{shock}} = \left [ \frac{(p_1 - p_0)(e_0 + p_1)}
{(e_1 - e_0)(e_1 + p_0)} \right]^{\frac{1}{2}}
\end{equation}
is the velocity of the shock front, $p_{ \rm 0}$ ($e_{ \rm 0}$) 
the pressure (energy density) in the shock front region and
$p_{ \rm 1}$ ($e_{ \rm 1}$) in the stationary medium itself.
Eq.\eqref{eq:v_shock} has the following limits: For
$p_0 >> p_1$ we obtain $v_{\rm{shock}} \approx 1$, whereas
for a small perturbation, $p_0 \approx p_1$, we get
$v_{\rm{shock}} \approx c_s$.

We employ the microscopic transport model BAMPS to investigate Mach
Cones with different strength of dissipations in the medium using a
jet moving in positive $z$-direction,
initialized at $t = 0$ fm/c at the position $z = -0.8$ fm.
The jet is treated as a massless particle with zero
spatial volume and zero transverse momentum, that is, $p_z = E_{\rm jet} = 200$
GeV and $v_{\rm jet} = 1$. The energy and momentum deposition to the
medium is realized via collisions with medium particles. In this scenario
we neglect the deflection of the jet and it can not be stopped by the medium;
its energy and momentum is set to its initial value after every collision.

All simulations are
realized within a static and uniform medium of massless Boltzmann
particles and $T = 400$ MeV. For this study we consider only binary scattering
processes with an isotropic cross section among the bulk particles.
To save computational
runtime we reduce our problem to two dimensions. Here we choose
the $xz$-plane and apply a periodic boundary condition in $y$-direction. 

%
\begin{figure}[th]
\includegraphics[width=\columnwidth]{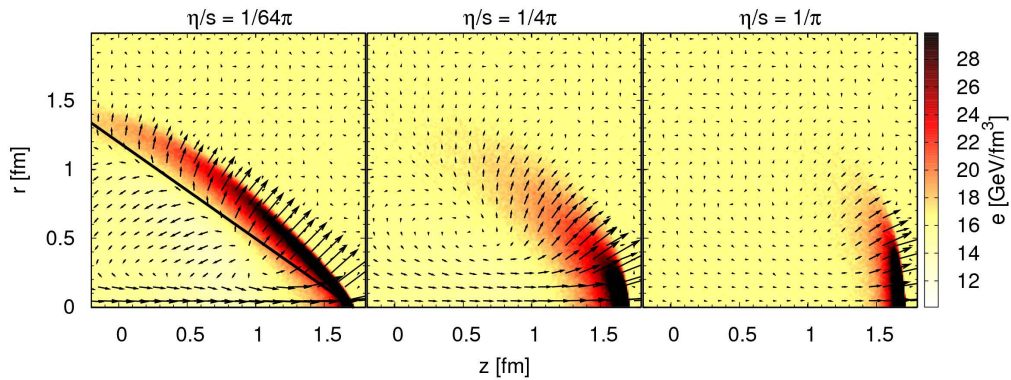}
\caption{(Color online) Scenario of a massless
jet with $p_z = E_{\rm jet} = 200$ GeV which can not be stopped
by the medium - the shape
of a Mach Cone shown for different viscosities of the medium,
$\eta/s = 1/64\pi$ (left), $\eta/s = 1/4\pi$
(middle), $\eta/s = 1/\pi$ (right). We show
the energy density plotted together with the velocity profile.
Additionally, in the left panel the linear ideal Mach Cone for
a very weak perturbation is shown by a solid line; its emission
angle is $\alpha_w = 54,73^\circ$.}
\label{fig:machCone}
\end{figure}
%

In Fig.\ref{fig:machCone} we demonstrate the transition from ideal
Mach Cone to a highly viscous one by adjusting the shear viscosity
over entropy density ratio in the medium from
$\eta/s = 1/64 \pi \approx 0.005$ to $1/\pi \approx 0.32$.
The energy deposition of the jet is approximately $dE/dx = 11 - 14$
GeV/fm. We show a snapshot at $t = 2.5$ fm/c.

Using an unphysical small viscosity of $\eta/s = 1/64\pi$ we
observe a strong collective behavior in form of a Mach Cone,
as shown in the left panel of Fig.\ref{fig:machCone}.
Due to the fact that the energy deposition is strong,
the shock propagates faster than the speed of sound through
the medium. For comparison, the ideal Mach Cone caused by a very weak
perturbation is given by a solid line with its emission angle
$\alpha_w = 54,73^\circ$.
Furthermore, a strong diffusion wake in direction of the jet,
characterized by decreased energy density, and a head shock in
the front are clearly visible.

%
\begin{figure}[th]
\includegraphics[width=\columnwidth]{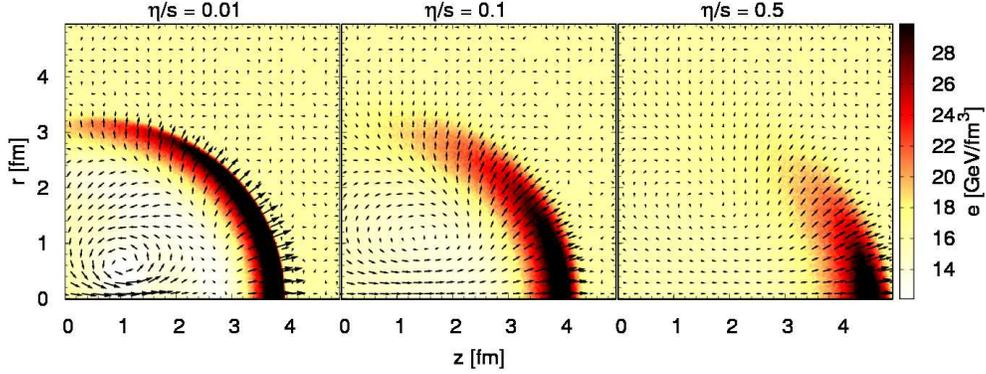}
\caption{(Color online) Scenarion of a deflectable jet with finite energy
$p_z = E_{\rm jet} = 20$ GeV - Induced Mach Cone structure
for different viscosities of the medium, $\eta/s = 0.01$ (left),
$\eta/s = 0.1$ (middle), $\eta/s = 0.5$ (right). We show the
energy density plotted together with the velocity profile.}
\label{fig:machCone_jetBundle}
\end{figure}
%

If we increase the viscosity of the medium to larger values, shown
in the middle and left panel of Fig.\ref{fig:machCone}, the
typical Mach Cone structure smears out and vanishes completely.
Due to stronger dissipation, the collective behavior gets weaker
because of less particle interactions in the medium with a larger
$\eta/s$. The results agree qualitatively with earlier studies
\cite{Bouras:2009nn,Bouras:2010hm},
where a smearing-out of the shock profile is observed with
higher viscosity.

In addition to the scenario of an unstoppable jet we demonstrate in
Fig.\ref{fig:machCone_jetBundle} the scenario of a massless jet
with finite energy which can also be deflected.
Its initial energy is set to $p_z = E_{\rm jet} = 20$ GeV, where the
starting point is $z = -0.3$ fm. We show the results for different
viscosities, $\eta/s \approx 0.01$ to $0.5$ at $t = 5.0$ fm/c.
In analogy to the results above we observe a clear Mach Cone
structure for small viscosities and a smearing out with larger values
of $\eta/s$. Only in the ideal case a strongly curved structure in which the
building up of a strong vortex is visible. The physical meaning
of these phenomena and also jets with the full pQCD cascade
have to be explored in future studies.

\section{Heavy quarks in BAMPS}

Initial heavy quark production during hard parton interactions in nucleon-nucleon scatterings and secondary production during the evolution of the QGP are studied. We use the event generator PYTHIA \cite{Sjostrand:2006za} to determine the initial heavy quark distributions, which agree with the experimental data from PHENIX \cite{Adare:2006hc_phenix_dsigmady}.
Nevertheless, these distributions have large uncertainties due to their sensitivity on the parton distribution functions in nucleons, the heavy quark masses as well as the renormalization and factorization scales (see \cite{Uphoff:2010sh} for a detailed analysis).
For the initial distribution of the gluonic medium we use three different approaches: the mini-jet model, a color glass condensate inspired model and also PYTHIA in combination with the Glauber model.

In the following we  give a brief overview of our results on heavy quark production in the QGP. More details concerning this section can be found in \cite{Uphoff:2010sh}.

\begin{figure}
\begin{minipage}[th]{0.49\textwidth}
\centering
\includegraphics[width=1.0\textwidth]{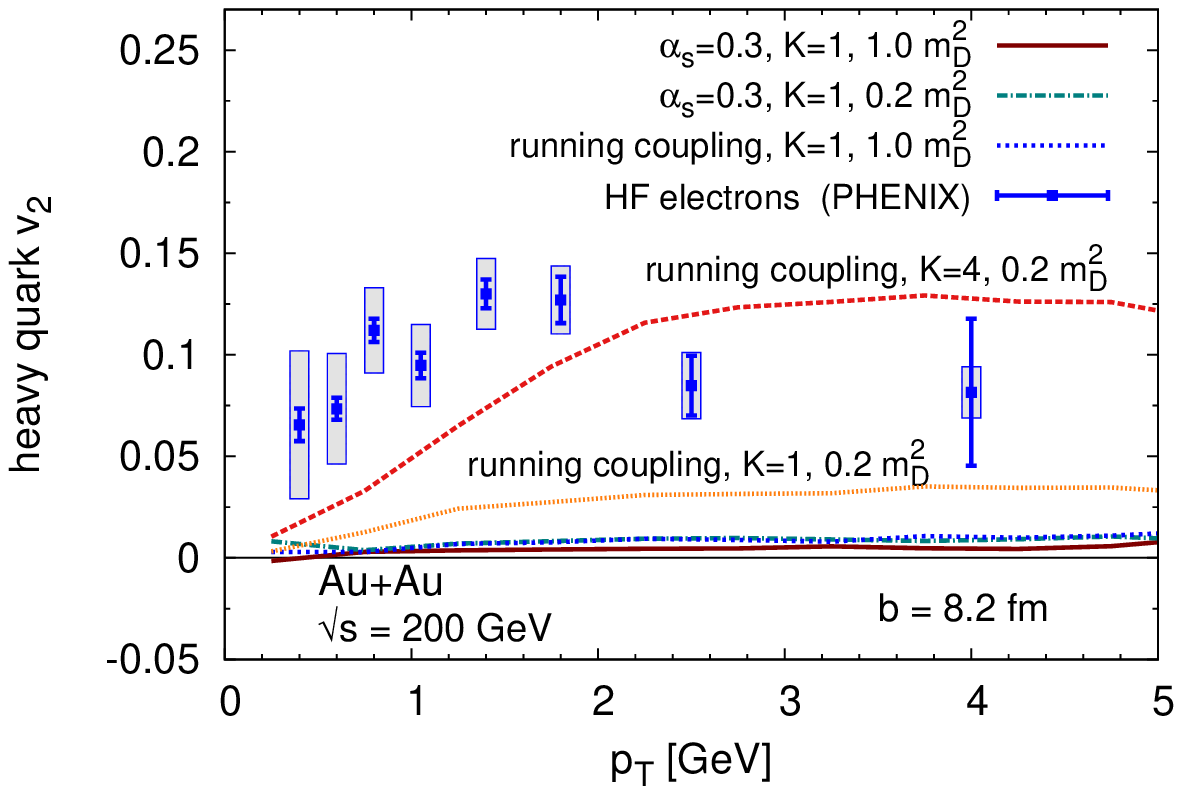}
\end{minipage}
\hfill
\begin{minipage}[th]{0.49\textwidth}
\centering
\includegraphics[width=1.0\textwidth]{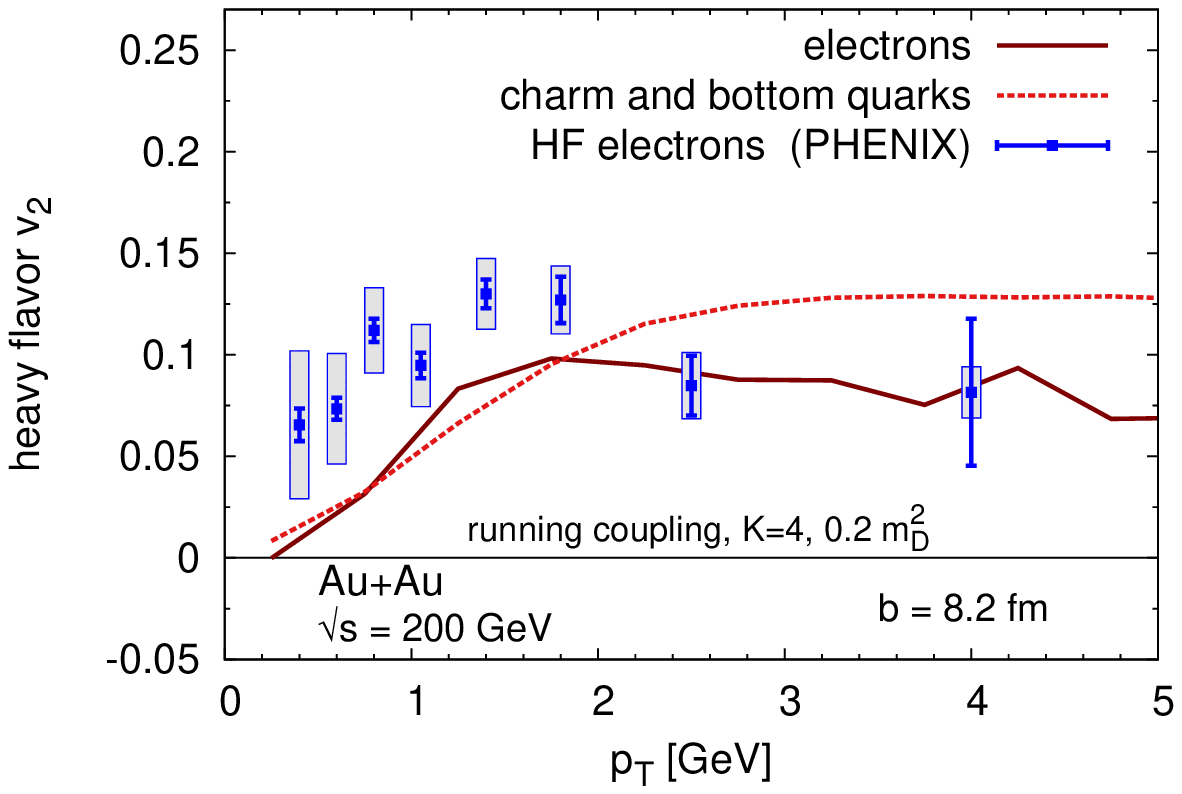}
\end{minipage}
\caption{Left panel: elliptic flow $v_2$ of heavy quarks with pseudo-rapidity $|\eta|<0.35$ at the end of the QGP phase for Au+Au collisions at RHIC with an impact parameter of $b=8.2 \, {\rm fm}$. For one curve the cross section of $gQ \rightarrow gQ$ is multiplied with a $K$ factor. For comparison, data of heavy flavor electrons \cite{Adare:2010de} is also shown. In contrast to the left panel, where the theoretical curves for $v_2$ on the quark level are plotted, the right panel depicts the flow of heavy quarks and electrons, which stem from the decay of $D$ and $B$ mesons produced by fragmentation.}
\label{fig:v2}
\end{figure}

Secondary heavy quark production in the QGP is studied within a full BAMPS simulation of Au+Au collisions at RHIC.
According to our calculations the charm quark production in the medium lies between 0.3 and 3.4 charm pairs, depending on the model of the initial gluon distribution, the charm mass and whether a $K=2$ factor for higher order corrections of the cross section is employed. However, compared to the initial yield these values are of the order of a few percent for the most probable scenarios. Therefore, one can conclude that charm production at RHIC in the QGP is nearly negligible. 

At LHC, however, the picture looks a bit different: Here the charm production in the QGP is a sizeable fraction of the initial yield and is even of the same order for some scenarios (with mini-jet initial conditions for gluons with a high energy density). In numbers, between 11 and 55 charm pairs are produced in the QGP.

Bottom production in the QGP, however is very small both at RHIC and LHC and can be safely neglected. As a consequence, all bottom quarks at these colliders are produced in initial hard parton scatterings.

The elliptic flow and the nuclear modification factor
\begin{align}
\label{elliptic_flow}
	v_2=\left\langle  \frac{p_x^2 -p_y^2}{p_T^2}\right\rangle \ , \qquad \qquad
R_{AA}=\frac{{\rm d}^{2}N_{AA}/{\rm d}p_{T}{\rm d}y}{N_{\rm bin} \, {\rm d}^{2}N_{pp}/{\rm d}p_{T}{\rm d}y}
\end{align} 
($p_x$ and $p_y$ are the momenta in $x$ and $y$ direction in respect to the reaction plane)
of heavy quarks at mid-rapidity are observables which are experimentally measurable and reflect the coupling of heavy quarks to the medium. A large elliptic flow comparable to that of light partons indicates a strong coupling to the medium. On the other hand a small $R_{AA}$ is a sign for a large energy loss of heavy quarks. Experimental results reveal that both quantities are on the same order as the respective values for light particles \cite{Adare:2010de,Abelev:2006db,Adare:2006nq}.

As we have recently shown \cite{Uphoff:2010fz} elastic scatterings of heavy quarks with the gluonic medium using a constant coupling $\alpha_s = 0.3$ and the Debye mass for screening the $t$ channel cannot reproduce the experimentally measured elliptic flow.  In order to explain the data one would need a $40-50$ times larger cross section than the leading order one. Of course, this $K$ factor is too large to represent the contribution of higher order corrections. However, as we demonstrated in Ref.~\cite{Uphoff:2010sy} and is shown in the following, the discrepancy with the data can be lowered -- even on the leading order level -- by a factor of 10 by taking the running of the coupling into account and by improving the incorporation of Debye screening. The remaining factor of 4 difference could then indeed stem from neglecting higher order effects, which, however, must be checked in a future project.

The following calculations are done analogously to \cite{Uphoff:2010sy,Gossiaux:2008jv,Peshier:2008bg}. An effective running coupling is obtained from measurements of $e^+e^-$ annihilation and non-strange hadronic decays of $\tau$ leptons \cite{Gossiaux:2008jv,Dokshitzer:1995qm}.

Since the $t$ channel of the $g Q \rightarrow g Q$ cross section is
divergent it is screened with a mass proportional to the Debye mass, which
is calculated by the common definition $m_{D}^2 = 4 \pi \, (1+N_f/6) \,
\alpha_s(t) \, T^2$ with the running coupling. The proportionality factor
$\kappa$ of screening mass and Debye mass is mostly set to 1 in the
literature without a sophisticated reason. However, one can fix this factor
to $\kappa = 0.2$ by
comparing the ${\rm d}E/{\rm d}x$ of the Born cross section with $\kappa$
to the energy loss within the hard thermal loop approach to $\kappa
\approx 0.2$ \cite{Gossiaux:2008jv,Peshier:2008bg}.

\begin{figure}
\begin{minipage}[th]{0.49\textwidth}
\centering
\includegraphics[width=1.0\textwidth]{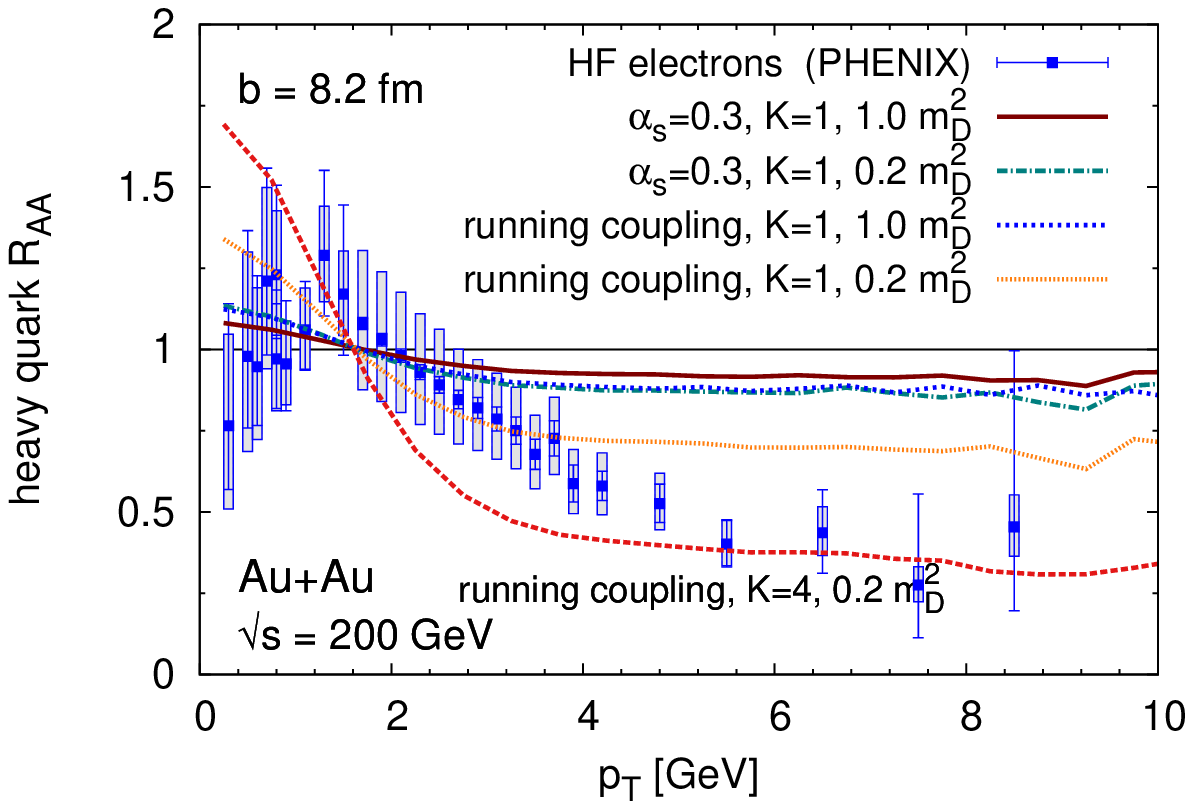}
\end{minipage}
\hfill
\begin{minipage}[th]{0.49\textwidth}
\centering
\includegraphics[width=1.0\textwidth]{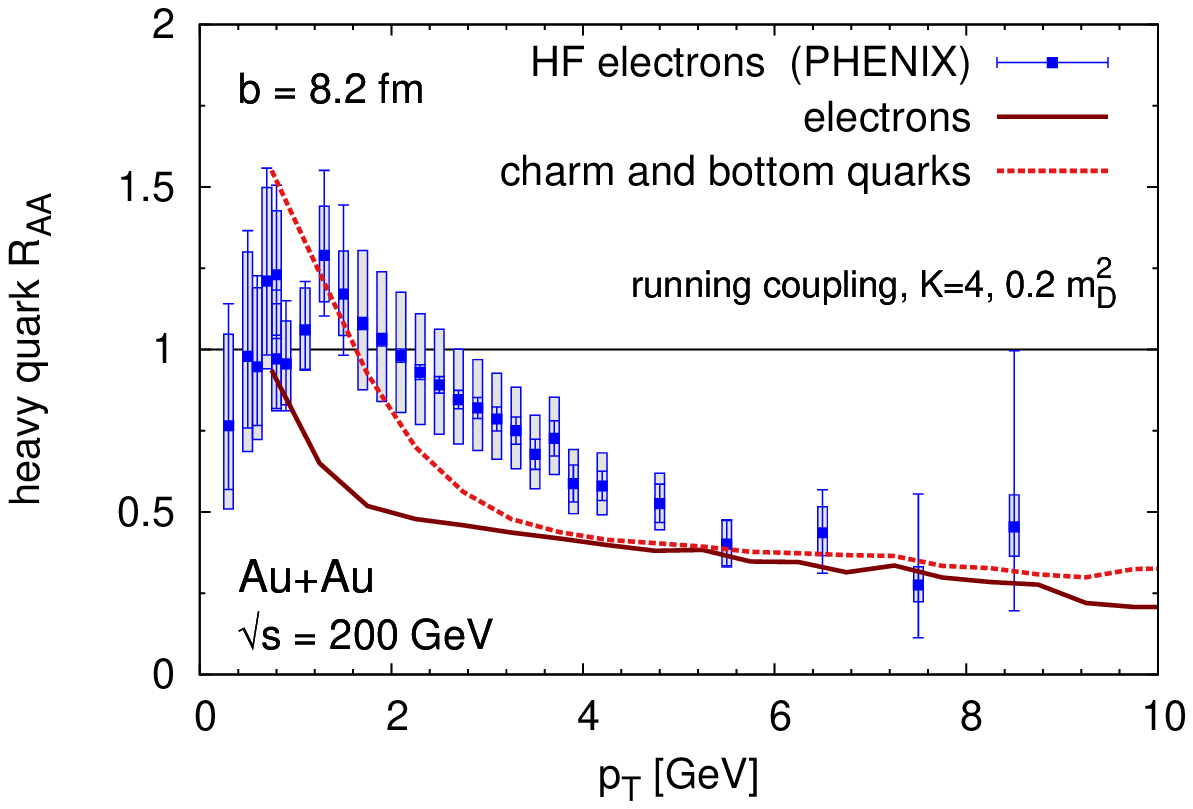}
\end{minipage}
\caption{As in Fig.~\ref{fig:v2}, but the nuclear modification factor $R_{AA}$ of heavy flavor is shown instead of $v_2$.}
\label{fig:raa}
\end{figure}

Fig.~\ref{fig:raa} depicts the $R_{AA}$ of heavy quarks, which shows for $K=4$ the same magnitude of suppression as the data.

These improvements lead to an enhanced cross section which also increases the elliptic flow. The left panel of Fig.~\ref{fig:v2} shows $v_2$ as a function of the transverse momentum $p_T$ for the leading order cross section without any improvements, with the running coupling, with the corrected Debye screening and with both modifications.

The elliptic flow of the latter reproduces the order of magnitude of the data, if the cross section is multiplied with $K=4$, which is much smaller than the previous employed $K=40-50$ and lies in a region which could account for higher order corrections. However, one has to check if these corrections have indeed a similar effect as a constant $K$ factor of 4. Therefore, the calculation of the next-to-leading order cross section is planned for the near future and will complement $2 \leftrightarrow 3$ interactions for gluons, which are already implemented in BAMPS \cite{Xu:2004mz}. The shapes of the theoretical curve and of the data points are, however, slightly different. This is an effect of hadronization and decay to electrons, which is not shown in the left panel of Fig.~\ref{fig:v2}. If one takes those two effects into account, the agreement of the data and the theoretically curve is much better, in particular for high $p_T$ (see right panel of Fig.~\ref{fig:v2}). We performed the fragmentation of charm (bottom) quarks to $D$ ($B$) mesons via Peterson fragmentation \cite{Peterson:1982ak} and used PYTHIA for the decay to electrons. At low $p_T$ the agreement between the theoretical curve and data becomes worse, since Peterson fragmentation is not the correct model here and coalescence may play a role.


\acknowledgments
The authors are grateful to the Center for Scientific 
Computing (CSC) at Frankfurt University for the computing resources.
I.\ B., J.\ U. and C.\ W. are grateful to Helmholtz Graduate School
for Hadron and Ion Research .

This work was supported by the Helmholtz International Center
for FAIR within the framework of the LOEWE program 
launched by the State of Hesse.


\end{document}